# Cooper pair trajectories in superconducting slab at self-field conditions

E.F. Talantsev[1,2*] and R.C. Mataira[3]

[1]M.N. Mikheev Institute of Metal Physics, Ural Branch, Russian Academy of Sciences,
18, S. Kovalevskoy St., Ekaterinburg, 620108, Russia

[2]NANOTECH Centre, Ural Federal University, 19 Mira St., Ekaterinburg, 620002, Russia

[3]Robinson Research Institute, Victoria University of Wellington, 69 Gracefield Road,
Lower Hutt, 5040, New Zealand

*E-mail: evgeny.talantsev@imp.uran.ru

**Abstract**

Dissipative-free electric current flow is one of the most fascinating and practically important property of superconductors. Theoretical consideration of the charge carriers flow in infinitely long rectangular slab of superconductor in the absence of external magnetic field (so called, self-field) is based on an assumption that the charge carriers have rectilinear trajectories in the direction of the current flow whereas the current density and magnetic flux density are decaying towards superconducting slab with London penetration depth as characteristic length. Here, we calculate charge particle trajectories (as single electron/hole, as Cooper pair) at self-field conditions and find that charge carriers do not follow intuitive rectilinear trajectories along the slab surface, but instead ones have meander shape trajectories which cross the whole thickness of the slab. Moreover, if the particle velocity is below some value, the charge moves in opposite direction to nominal current flow. This disturbance of the canonical magnetic flux density distribution and backward movement of Cooper pairs can be entire mechanism for power dissipation in superconductors.

**Cooper pair trajectories in superconducting slab at self-field conditions**

1. **Introduction**

Transport current flow and self-generating magnetic field in superconducting slab have been under theoretical considerations since famous London brothers' paper [1] over decades [2-19]. The understanding of superconducting current flow in rectangular slab has direct impact on superconducting technology, because second generation high-temperature superconductors have a design in form of thin superconducting film deposited on a metallic substrate with shunting layers on both sides of the tape [20-22].

London and London [1] derived an equation for the distribution of dissipative-free transport current density, $J$, across infinite wide superconducting slab with thickness of $2b$:

$$J_y(z) = J_{y,surf} \cdot \frac{\cosh\left(\frac{z}{\lambda}\right)}{\cosh\left(\frac{b}{\lambda}\right)} \quad (1)$$

where $\lambda$ is the London penetration depth, $2b$ is the slab thickness, $J_{surf}$ is the surface current density (here we use Cartesian coordinates, such that **X** is oriented in the direction of **B**, **Y** is oriented in the direction of the current **I** and **Z** is oriented in normal direction to the surface of the slab). Magnetic flux density, $B_x(z)$, has $z$-axis dependence [1]:

$$B_x(z) = |B_{x,surf}| \cdot \frac{\sinh\left(\frac{z}{\lambda}\right)}{\sinh\left(\frac{b}{\lambda}\right)} \quad (2)$$

Problem geometry is shown in Fig. 1.

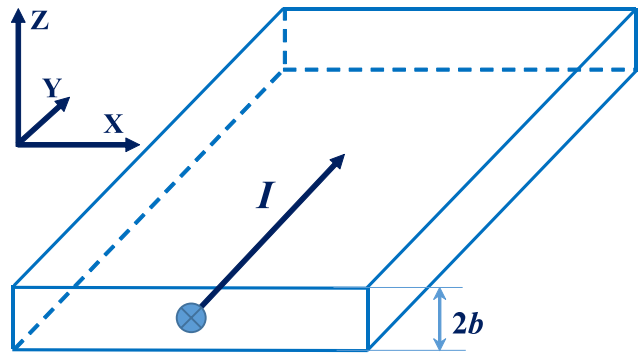

**Figure 1.** Geometry of the experiment.

Schematic representation of the transport current density, $J_y(z)$, across the thickness of the slab is shown in Fig. 2,a (for which exact distribution is given by Eq. 1). Magnetic flux density, $B_x(z)$, which describes by Eq. 2, is schematically shown in Fig. 2,b.

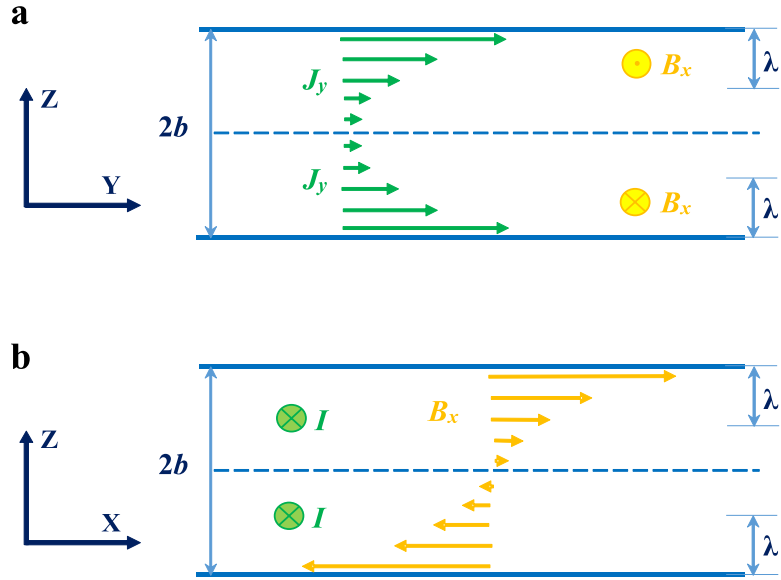

**Figure 2.** Transport current density, $J$ (a), and magnetic flux density, $B$ (b), distributions in the superconducting slab at self-field condition.

In this paper, we fix magnetic flux density, $B_x(z)$, distribution to be described by Eq. 2 and calculate Cooper pair trajectories in superconducting slab. Despite a fact that all our findings have general implication, in this paper, to be close to experimental data of the self-field critical current in real superconductors, we demonstrate all calculations for Cooper pairs which travel in the central cross-sectional plane of one of niobium thin films with thickness of $2b$ = 53 nm, width of $2a$ = 2.5 μm, and $J_c(sf, T = 2K) = 2.64 \cdot 10^{11}\ A/m^2$, reported by Rusanov *et al* [23].

2. **Basic equations**

To calculate Cooper pair trajectories, we employ numerical calculations by Euler's method, which (for axes geometry described in Fig. 1) utilises following equations:

$$\frac{d^2x}{dt^2} = 0 \tag{3}$$

$$\frac{d^2y}{dt^2} = \frac{q}{m^*} \cdot B_x(z) \cdot \frac{dz}{dt} \tag{4}$$

$$\frac{d^2z}{dt^2} = -\frac{q}{m^*} \cdot B_x(z) \cdot \frac{dy}{dt} \tag{5}$$

where $q$ and $m^*$ are the charge and the mass of the particle (details can be found elsewhere [24,25]).

To use these equations in a conjunction with Eq. 2, there is a need to compute:

1. $B_{x,surf}(z = \pm b)$, to calculate $B_x(z)$ by Eq. 2;

2. London penetration depth, $\lambda(T)$, which is used to calculate $B_x(z)$ by Eq. 2;

3. $v_y(t = 0) = \frac{dy}{dt}$, which is initial Cooper pairs velocity along Y-axis which will be calculated, while $v_x(t = 0) = v_z(t = 0) = 0$.

All necessarily values will be computed in the next section for particular niobium sample ($2b$ = 53 nm, width $2a$ = 2.5 μm, and $J_c(sf, T = 2K) = 2.64 \cdot 10^{11} \ A/m^2$) at the lowest available in experiment temperature of 2 K [23] and at the condition of the start of the electric power dissipation, which is designated by the critical current $I_c$ or critical current density, $J_c = I_c/(4 \cdot a \cdot b)$.

## 3. Results

To calculate $B_{x,surf}(z = \pm b)$ (Eq. 2) in given niobium film we use Ampere's law:

$$B_{x,surf}(z = \pm b) = \mp \mu_0 \cdot b \cdot J_c = \mp 8.79 \ mT \tag{6}$$

where $b$ and $J_c(sf, T = 2K) = 2.64 \cdot 10^{11} \ A/m^2$ are reported by Rusanov *et al.* [21] (more details can be found elsewhere [26]).

To deduce the London penetration depth, $\lambda(T = 2K)$, universal equation for isotropic superconductor [26,27]):

$$J_c(sf, T) = \frac{\phi_0}{4\cdot\pi\cdot\mu_0} \cdot \frac{\ln(1+\sqrt{2}\cdot\kappa(T))}{\lambda^3(T)} \cdot \left[\frac{\lambda(T)}{a}\cdot tanh\left(\frac{a}{\lambda(T)}\right) + \frac{\lambda(T)}{b}\cdot tanh\left(\frac{b}{\lambda(T)}\right)\right] \quad (7)$$

where $\kappa(T=2\ K) = \frac{\lambda(T=2\ K)}{\xi(T=2\ K)} = 1.0$ for niobium [28] and $\xi(T)$ is the coherence length, was numerically solved. Computed value is $\lambda(T=2K) = 79\ nm$.

To calculate trajectories there is a need to determine Cooper pair velocity, $v_y(t=0) = v_{C,c}$, at the condition of critical current flow. The value of $v_{C,c}$ is determined by the use of the momentum of the Cooper pair expressed through its de Broglie wavelength, $\lambda_{dB,c}(T)$ [27], at critical current:

$$p_C(T) = m^* \cdot v_{C,c}(T) = \frac{h}{\lambda_{dB,c}(T)} = \frac{h}{\frac{4\cdot\lambda(T)}{\ln\left(1+\sqrt{2}\cdot\frac{\lambda(T)}{\xi(T)}\right)}} \quad (8)$$

$$v_{c,c}(T) = \frac{h}{m^*} \cdot \frac{\ln\left(1+\sqrt{2}\cdot\frac{\lambda(T)}{\xi(T)}\right)}{4\cdot\lambda(T)} \quad (9)$$

Substituting deduced $\lambda(T=2K) = 79\ nm$ and $m^* = m^*_{eff} \cdot (2 \cdot m_e)$, where $m^*_{eff} = 1.3$ is taken from Ref. 29, results in:

$$v_y(t=0) = v_{C,c}(T=2\ K) = 7.8 \cdot 10^2\ \frac{m}{s} \quad (10)$$

Based on experimental fact that practically all superconductors are hole type conductors [30], we consider the Cooper pair charge $q = +2e$ in our calculations (Eqs. 3-5).

Calculated trajectory for the Cooper pair, for which the starting point is at the surface of the films, i.e. (0, 0, b), and initial velocity is $(0, v_{C,c}, 0)$, is shown in Fig. 3. Unexpectedly, the trajectory has meander shape and the Cooper pair is oscillating from one surface of the film to another while its travel to the current direction. The same meandering type of trajectory is found for others initial position of the Cooper pair (0, 0, z), except particles which start their movement from the origin (0, 0, 0). To demonstrate this, in Fig. 3 we also show the trajectory for the Cooper pair with starting point at (0, 0, -b/2).

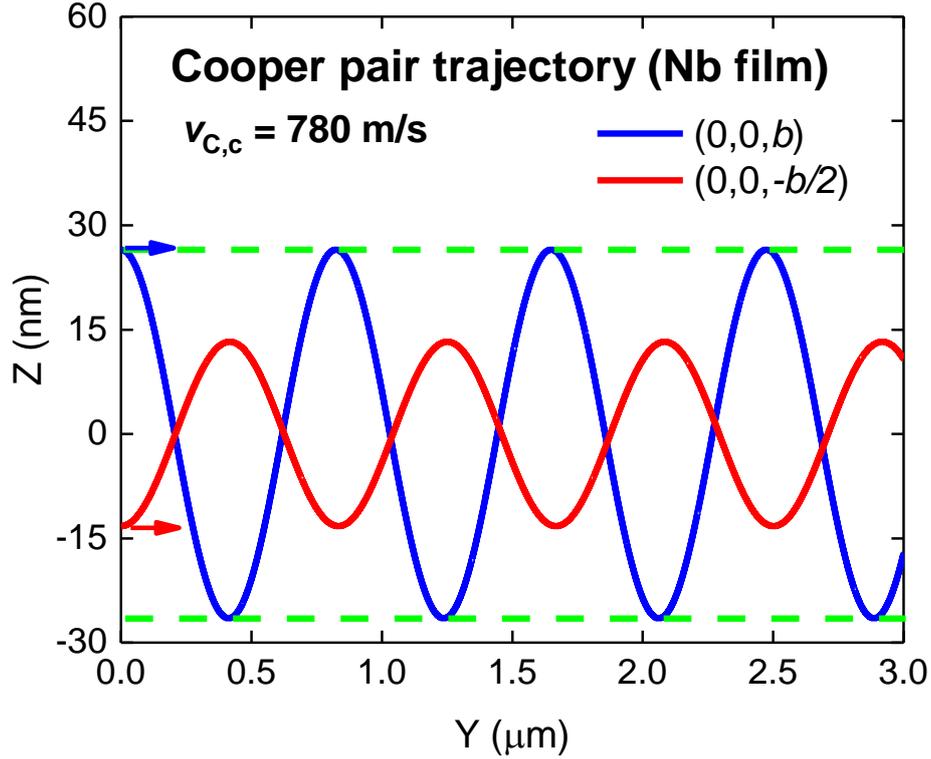

**Figure 3.** Cooper pair trajectories in thin niobium film ($2b = 53$ nm; $J = 2.64 \cdot 10^{11}$ $A/m^2$) with $v_y(t = 0) = 7.8 \cdot 10^2$ $m/s$ and two starting positions: (0,0,*b*) (blue line) and (0,0,-*b*/2) (red line). Green dash lines indicate thin film surfaces. Initial velocity vectors for both particles are shown for clarity.

It can be seen (Fig. 3) that in presumption that London and London [1] equation for magnetic flux density distribution (Eq. 2) is true, Cooper pair trajectories do not follow intuitive assumption that ones are traveling within rectilinear trajectories, as ones are schematically shown in Fig. 2 (which can be also found in many textbooks on superconductivity, see, for instance [29]).

We also studied the influence of the Cooper pair initial velocity, $v_C$, on the trajectory. The most remarkable findings are represented in Figs. 4,5. In Fig. 4 we show the trajectories for $v_C = 4 \cdot v_{C,c}$ (for initial particle position at (0, 0, *b*)) and $v_C = \frac{v_{C,c}}{4}$ (for initial particle position at (0, 0, -*b*/2). It can be seen that meander-type trajectories are remaining and the difference is belonging the change in spatial period. Total direction of the particle movement is along initial direction of $v_C$ velocity.

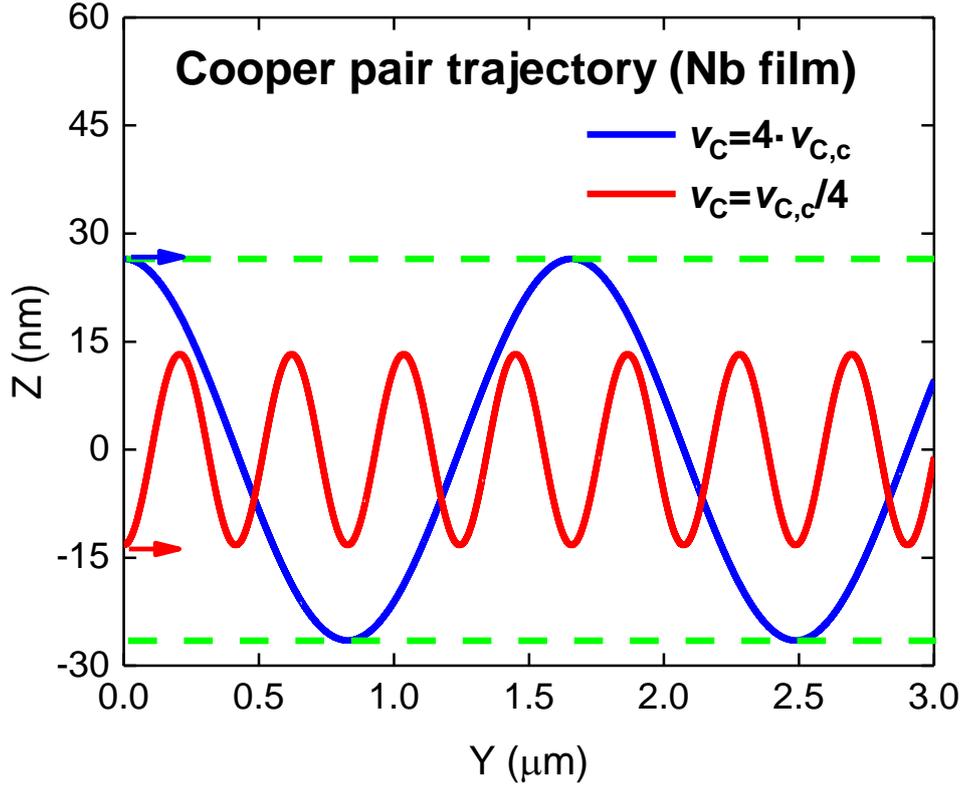

**Figure 4.** Cooper pair trajectories in thin niobium film ($2b = 53$ nm; $J = 2.64 \cdot 10^{11}\ A/m^2$) with $v_y(t=0) = 4 \cdot v_{C,c}$ (blue line) and $v_y(t=0) = v_C = \frac{v_{C,c}}{4}$ (red line). Green dash lines indicate thin film surfaces. Initial velocity vectors for both particles are shown for clarity.

However, while $v_y(t=0) = v_C$ is significantly decreased in comparison with $v_{C,c}$, a new puzzling feature is appeared. This feature is shown in Fig. 5 where two Cooper pairs start to move with the same initial velocity of $v_y(t=0) = \frac{v_{C,c}}{90}$, however, as in both previous calculations (Figs. 3,4) particles are placed at different positions, i.e. (0, 0, *b*) and (0, 0, -*b*/2). For both particles, initial velocities are shown by arrows for clarity. The particle starts its movement at position of (0, 0, *b*) effectively transfers the electric charge in opposite direction to nominal current flow (Fig. 5). However, the particle which is located closer to the slab centre is still transfer the electric charge along nominal direction (Fig. 5).

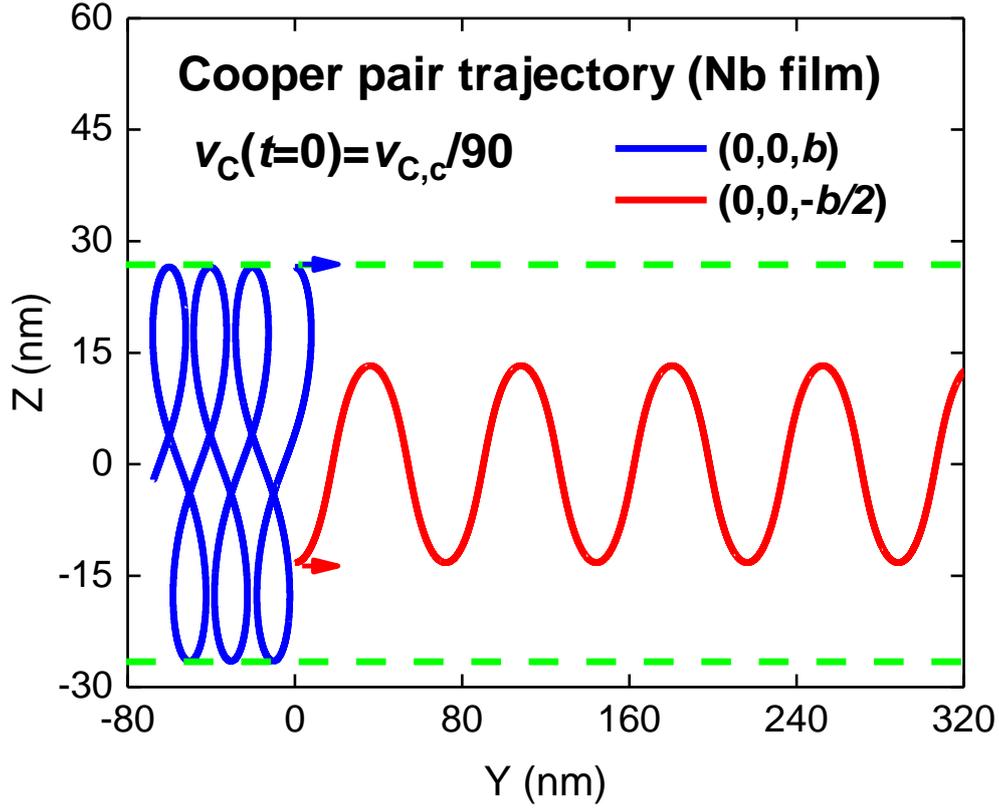

**Figure 5.** Cooper pair trajectories in thin niobium film ($2b = 53$ nm; $J = 2.64 \cdot 10^{11}\ A/m^2$) with $v_y(t=0) = v_C = \frac{v_{C,c}}{90}$ for two particles with initial positions at $(0, 0, b)$ (blue line) and $(0, 0, -b/2)$ (red line). Green dash lines indicate thin film surfaces. Initial velocity vectors for both particles are shown for clarity.

This means that superconducting slab is splatted in three layers: two surface layers where current is flowing in opposite direction (to the nominal direction of the current flow), and in internal layer, where electric current is still flowing in the direction of the nominal current flow.

This splitting is transformed into spatial split, if the velocity becomes lower. In Fig. 6 we show trajectories for the initial velocity of $v_y(t=0) = \frac{v_{C,c}}{150}$, for which trajectories of two particles are completely splatted by spatial separation.

The appearance of Cooper pairs which exhibit much lower velocity than $v_{C,c}$ can be originated by the pair scattering on lattice disturbances. Because these particles (at some conditions) have backward movement, this can be entire mechanism for the electric power dissipation, which, however, does not necessarily mean, that the pair should decay.

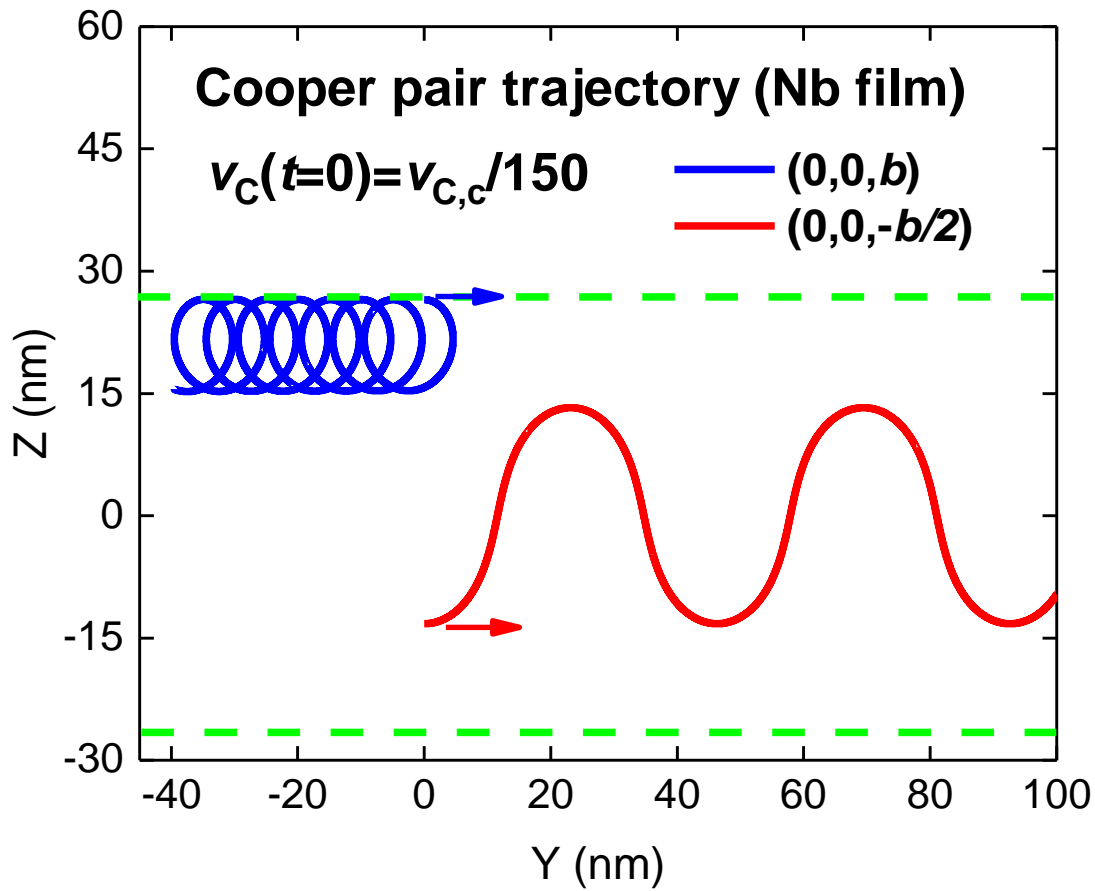

**Figure 6.** Cooper pair trajectories in thin niobium film ($2b = 53$ nm; $J = 2.64 \cdot 10^{11}\ A/m^2$) with $v_y(t = 0) = v_C = \frac{v_{C,c}}{150}$ for two particles with initial positions at $(0, 0, b)$ (blue line) and $(0, 0, -b/2)$ (red line). Green dash lines indicate thin film surfaces. Initial velocity vectors for both particles are shown for clarity.

## 4. Conclusion

Our findings can be summarized in three general conclusions:

1. Cooper pair trajectories in superconducting slab at self-field conditions are not follow intuitive rectilinear lines with reducing density of such moved Cooper pairs by the law:

$$J_y(z) = J_{y,surf} \cdot \frac{cosh\left(\frac{z}{\lambda}\right)}{cosh\left(\frac{b}{\lambda}\right)}$$

but instead the trajectories have meander-type shape, which cover the whole cross-section of the superconducting slab.

2. Depends on Copper pair velocity, the meander-type trajectories can have backwards direction to the nominal current direction. Copper pairs travels at surface layer are more likely to be affected to this backward movement.

3. This backward movement can create the appearance of the electric field along the slab in the direction of the current flow, which is effectively is the power dissipation. However, this mechanism does not require that Cooper pairs should be decayed.

**Acknowledgement**

EFT thanks financial support provided by the state assignment of Minobrnauki of Russia (theme "Pressure" No. AAAA-A18-118020190104-3) and by Act 211 Government of the Russian Federation, contract No. 02.A03.21.0006.